# *Identifying the Structural Basis for the Increased Stability of the Solid Electrolyte Interphase Formed on Silicon with the Additive Fluoroethylene Carbonate*

Yanting Jin,[†] Nis-Julian H. Kneusels,[†] Pieter C. M. M. Magusin,[†] Gunwoo Kim,[†,‡] Elizabeth Castillo-Martínez,[†] Lauren E. Marbella,[†] Rachel N. Kerber,[†] Duncan J. Howe,[†] Subhradip Paul,[§] Tao Liu,[†] and Clare P. Grey*,[†]

[†]Department of Chemistry, University of Cambridge, Lensfield Road, Cambridge CB2 1EW, United Kingdom

[‡]Cambridge Graphene Centre, University of Cambridge, Cambridge, United Kingdom CB3 0FA

[§]DNP MAS NMR Facility, Sir Peter Mansfield Magnetic Resonance Centre, University of Nottingham, Nottingham NG7 2RD, United Kingdom

**ABSTRACT:** To elucidate the role of fluoroethylene carbonate (FEC) as an additive in the standard carbonate-based electrolyte for Li-ion batteries, the solid electrolyte interphase (SEI) formed during electrochemical cycling on silicon anodes was analyzed with a combination of solution and solid-state NMR techniques, including dynamic nuclear polarization. To facilitate characterization via 1D and 2D NMR, we synthesized $^{13}$C-enriched FEC, ultimately allowing a detailed structural assignment of the organic SEI. We find that the soluble PEO-like linear oligomeric electrolyte breakdown products that are observed after cycling in the standard ethylene carbonate (EC)-based electrolyte are suppressed in the presence of 10 vol % FEC additive. FEC is first defluorinated to form soluble vinylene carbonate and vinoxyl species, which react to form both soluble and insoluble branched ethylene-oxide based polymers. No evidence for branched polymers are observed in the absence of FEC.

## INTRODUCTION

The formation of a solid electrolyte interphase (SEI) on the electrode surface is critical for the cycle life of lithium ion batteries (LIBs). Stable SEI formation on the anode is particularly important, because current LIBs operate outside the stable voltage window of the organic carbonate based liquid electrolyte (<1.3 - 0.8 V vs. Li).[1] Without such a SEI, continuous breakdown of the electrolyte proceeds uninhibited upon further cycling, which irreversibly consumes the Li source and leads to capacity fading.[2,3] Although electrolytes have been optimized so that they form a stable SEI on the commercial graphite anode, preventing significant capacity fade even after extended cycling, problems remain for next generation, higher-capacity anodes such as Si, Sn and Li metal.[4] Optimizing the SEI that forms on Si is particularly important because Si anodes offer a theoretical capacity (3579 mAh/g) that is nearly an order of magnitude greater than that of graphite (372 mAh/g).[4] Unfortunately, the lithiation and delithiation of Si electrodes is accompanied by a large volume expansion (~300%, for the formation of Li$_{3.75}$Si)[5], which leads to continuous exposure of fresh Si surface during cycling, and results in uncontrolled SEI growth and poor cycling performance.[6] The mechanisms and reactions involved in SEI formation on Si differ from those on the graphite, because the soluble and insoluble decomposition products can con tinue to react largely unimpeded at lower voltages (due to exposed fresh lithium silicide surfaces) and over multiple cycles. In an effort to stabilize the SEI on Si, sacrificial electrolyte additives are commonly used to improve the cycle life.[8] Among the numerous possible electrolyte additives, fluoroethylene carbonate (FEC) is the most widely used one and has been shown to improve the capacity retention of Si[9], Li anodes[10] and a series of cathode materials.[11,12] However, the mechanisms by which these additives modify the nature of the SEI are not well understood due to the inherent difficulties associated with characterizing disordered interfacial structures. Characterization of the few nanometer-thick SEI layer is further convoluted by the fact that the material is usually air-sensitive and amorphous, the former presenting challenges in sample preparation and the latter with structural assignment with, for example, diffraction-based techniques. Despite these difficulties, electrolyte breakdown products in the presence of FEC have been characterized by a variety of spectroscopic and modeling approaches to provide insight into the chemical composition of the SEI on Si. For instance, Fourier Transform infrared (FTIR) spectroscopy suggested that FEC likely transforms to vinylene carbonate (VC) through HF elimination and subsequently polymerizes to form poly(VC)-type species.[13] In contrast, Nakai et al. studied the reduction of EC-free, FEC-based electrolyte on Si by X-ray photoelectron spectroscopy (XPS) and flight–secondary ion mass spectrometry (ToF-SIMS) and found evidence for oxygen-deficient species such as polyene.[11] The presence of low-oxygen content polyenes was further supported by energy-dispersive X-ray spectroscopy (EDS)[12] and on-line electrochemical mass spectrometry (OEMS)[13] performed during the reduction of FEC on Si. An FEC decomposition intermediate, the vinoxyl radical (HC(=O)CH$_2$•), was proposed by Balbuena and co-workers[17] through ab initio calculations, and was later experimentally detected by Shkrob et al. using electron paramagnetic resonance spectroscopy.[18] The vinoxyl radical is formed via a one-electron reduction of FEC, resulting in the formation of LiF and the concurrent loss of $CO_2$[17,18]. The vinoxyl radical can then react further with EC or FEC to form oligomers of the form HC(=O)CH$_2$OR[17,18] and is believed to initiate the polymerization that eventually results in a highly cross-linked network.[18]

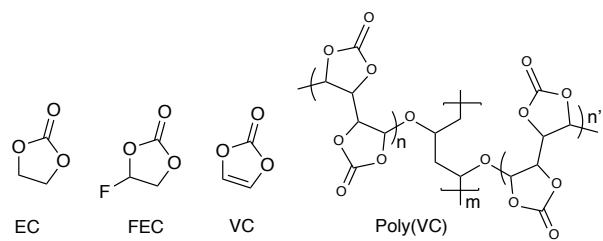

**Scheme 1.** Structures of ethylene carbonate (EC), fluoroethylene carbonate (FEC), vinylene carbonate (VC), and a possible structure of the polymer "poly(VC)" as reported in reference [7].

During cycling, the Si surface can be modified by reacting with electrolyte. XPS study by Philippe et al. suggested that Si can be oxidized during cycling, forming lithium silicate and fluorosilicate in the presence of $LiPF_6$.[16] Nakai et al. suggested that FEC can protect the Si surface against electrolyte oxidation, probably due to the formation of a passivation layer on Si surface.[14] A recent detailed XPS study by Schroder et al. concluded that the reduction of FEC leads to the formation of a kinetically stable SEI comprising a more dense layer of lithium fluoride and lithium oxide near the Si surface, which improve lithiation kinetics.[20]

Michan et al. used solid-state NMR (ssNMR) to analyze the precipitate formed by chemically reducing FEC with lithium naphthalenide and proposed a structure for the resulting polymeric species (Scheme 1, poly(VC)).[7] A $^{13}$C NMR resonance at 100 ppm, which had not been previously observed in the decomposition products of EC, was assigned to an acetal carbon (a protonated carbon environment adjacent to two oxygen groups), and represents a potential cross-linking unit in poly(VC). Of note, acetal carbons were observed in pioneering NMR studies of the SEI formed on graphite by Leifer et al.[21] However, some of the resonances in Michan's study were partially obscured by the presence of excess lithium naphthalenide in the precipitate which complicated further analysis. Perhaps more importantly, the chemically induced FEC-polymer may differ from the organic SEI generated through electrochemical reduction, necessitating further studies on SEI decomposition products formed in battery materials over the course of cycling.

Here, we use a combination of solution and solid-state NMR to characterize and rationalize the formation of both the soluble and insoluble FEC decomposition products in the SEI on silicon nanowires (SiNWs). The SiNWs were chosen as they can be studied in the absence of binder to simplify the analysis. SiNWs were subjected to a constant current for long-term cycling to electrochemically generate sufficient SEI for structural characterization. In order to overcome the inherently low sensitivity of NMR, we synthesized uniformly labeled $^{13}C_3$ FEC and performed ssNMR experiments with dynamic nuclear polarization (DNP) enhancement to provide a more comprehensive understanding of the organic SEI. Our results clearly show that chemically distinct intermediates are formed in both the soluble and insoluble SEI products upon addition of FEC that may be responsible for increased stability in the presence of this additive.

## EXPERIMENTAL METHODS

**Synthesis of $^{13}C_3$-fluoroethylene carbonate.** $^{13}C_3$ FEC was synthesized from $^{13}C_3$ EC via chlorination of EC and subsequent fluorination using standard Schlenk techniques under N$_2$-atmosphere. $^{13}C_3$ EC (200 mg, 2.27 mmol, 1.00 eq.) was suspended in 5 mL of carbon tetrachloride. Sulfuryl chloride (0.19 mL, 2.38 mmol, 1.05 eq.) and azobisisobutyronitrile (AIBN) (15 mg, 0.91 mmol, 0.04 eq.) were added and the further AIBN was added twice every 30 minutes. The reaction mixture was left stirring at 65 °C for 16 h before the solvent was removed in vacuo and the liquid residue was purified by column purification (silica 100:1, dichloromethane, R$_f$ = 0.50). $^{13}C_3$-Chloroethylene carbonate (180 mg, 1.47 mmol, 65%) was received as a clear colorless liquid.

Anhydrous potassium fluoride (2.50 g, 43.0 mmol, 29.3 eq.) was suspended in 7 mL of acetonitrile abs. and $^{13}C_3$-chloroethylene carbonate (180 mg, 1.47 mmol, 1.00 eq.), dissolved in 3 mL acetonitrile abs., was added. The mixture was stirred at 75 °C for 16 h, filtered, and the solid residue was washed with 10 mL of acetonitrile. The filtrates were combined and the solvent was removed in vacuo. The dark liquid was purified by column purification (silica 100:1, dichloromethane, R$_f$ = 0.45). $^{13}C_3$ FEC (40 mg, 0.38 mmol, 26 %) was separated as a clear colorless liquid.

**Synthesis of silicon nanowires.** Silicon nanowires were synthesized by chemical vapor deposition (CVD) as previously described[22]. Briefly, 50 nm gold was thermal-sputtered onto 20 μm-thick stainless steel (SUS304, Agar Scientific) foil. The gold-coated stainless steel foil was cut into 1×1 cm$^2$ substrates, which were transferred into the CVD chamber. The substrates were heated in 1 mBar of argon atmosphere at 510-530 °C for 10 min to anneal the gold catalyst. After annealing, a mixture of argon and silane gases (Ar / SiH$_4$= 100 sccm / 20 sccm) was introduced. The growth was carried out in 15 mBar at 510 – 530 ˚C for 15 min. The substrates were weighed before and after CVD growth to determine the mass of SiNWs. The average mass loading of SiNWs was around 0.5-0.8 mg/cm$^2$.

**Electrolyte preparation and coin cell assembly.** The five different electrolyte formulations used here are listed in Table 1. The LP30 + $^{13}C_3$ EC enrichment of electrolytes was prepared by mixing $^{13}C_3$ EC with a non-labeled EC/DMC in a 1:1:2 v/v/v ratio, then dissolving the LiPF$_6$ salt into the solvent to achieve a final concentration of 1 M. The LP30 + FEC and LP30 + $^{13}C_3$ FEC electrolytes were prepared by either adding 0.5 mL of FEC or $^{13}C_3$ FEC into 5 ml of commercial LP30 electrolyte. All the electrolytes were stored in aluminum bottles.

**Table 1. Electrolyte formulation with 1M LiPF$_6$ in different solvent mixtures**

| Electrolyte solvents | Abbreviation |
| --- | --- |
| EC/DMC = 50/50 (v/v) | LP30 |
| $^{13}C_3$ EC/EC/DMC = 25/25/50 (v/v/v) | LP30 + $^{13}C_3$ EC |
| EC/DMC/FEC = 50/50/10 (v/v/v) | LP30 + FEC |
| EC/DMC/$^{13}C_3$ FEC = 50/50/10 (v/v/v) | LP30 + $^{13}C_3$ FEC |
| $^{13}C_3$ EC/EC/DMC/FEC=25/25/50/10 (v/v/v/v) | LP30 + $^{13}C_3$ EC + FEC |

SiNWs electrodes were then assembled into Li-half 2032 coin cells using the five electrolytes. Porous glass fiber mats (Whatman GF/B, B 1 mm thick) were used as separators and around 10 drops (~ 0.2 mL) of electrolyte were used for each cell. All the assembling procedures were carried out in an Ar filled glovebox (H$_2$O < 0.1ppm, O$_2$ < 0.1ppm). The coin cells were discharged/charged at constant current (C/30, 120 mA/g) between

0.001 V – 2 V regions for 1st and 30th cycles using a Biologic VSP or MPG-2. Approximately 50 days were needed to complete 30 cycles. The slow cycling protocol ensures that the electrolyte solvents are being held at low voltage for sufficiently long time for extensive SEI formation. The electrochemical results obtained for the enriched electrolyte were similar to those of the non-enriched electrolyte: the cycling performance is mainly influenced by the presence of FEC.

**Solution NMR.** After the SiNW coin cells finished the 1st and 30th cycles, the cells were disassembled in an Ar filled glovebox. The glass fiber separators were extracted and soaked in 0.75 mL DMSO-$d_6$ for 2-3 min. The solution was then transferred to an airtight J-Young tube. Spectra were recorded on a 500 MHz Bruker Avance III HD, with a DCH (carbon observe) cryoprobe or Bruker AVANCE 400 equipped with a BBO probe. Detailed information about the pulse program can be found in SI. $^1$H and $^{13}$C NMR spectra were internally referenced to DMSO-$d_6$ at 2.50 ppm and 39.51 ppm, respectively.

**Solid-state NMR.** After cell disassembling, the SiNW electrodes were dried under vacuum overnight (~16-20 h) to remove the DMC and FEC; this procedure also removes most of the EC.[23] Note that the electrodes were not rinsed. After drying, the SiNW electrodes were scratched from the substrate and packed into rotors for ex-situ multinuclear ssNMR measurement.

$^1$H-$^{13}$C cross polarization (CP) of LP30 +$^{13}$C$_3$ EC sample was performed on a Bruker Avance III 700 (16.4 T) spectrometer using a 3.2 mm HXY probe at MAS frequency of 20 kHz, with CP contact time of 1 ms. RF nutation frequency were ($^1$H) 92.5 kHz (50 – 100 % linearly ramped during CP[24]), ($^{13}$C) 82.5 kHz, and SPINAL-64[25] $^1$H decoupling at 80 kHz. 3482 scans separated by a 3 s recycle interval were acquired over 3 h. The LP30 +$^{13}$C$_3$ FEC sample was measured on a Bruker Avance III HD 500 (11.7 T) spectrometer using a 2.5 mm HX probe at MAS frequency of 10 kHz, with CP contact time of 2 ms and SPINAL-64 $^1$H decoupling at 80 kHz. 24576 scans separated by a 3 s recycle delay were acquired over 20.5 h. The experimental parameters are summarized in Table S2. $^1$H and $^{13}$C, shifts were externally referenced to adamantane at 1.87 and 38.6 ppm (of CH$_2$ group), respectively.

**DNP NMR.** The cycled SiNW samples were sealed under Ar and transferred to the Nottingham DNP MAS NMR Facility in three layers of sealed plastic bag. The samples were then quickly pumped into the N$_2$-filled glovebox. Since the SiNWs were in the delithiated state, we assumed that there was no reaction between N$_2$ and the electrodes material. The sample was diluted with pre-dried KBr power by mixing homogeneously in a mortar. Then, a minimum amount of radical solution (4 mM TEKPol in 1,2-dichlorobenzene, DCB)[26,27] was added to wet the powder. The resulting paste-like samples were packed into the center of 3.2 mm sapphire MAS rotors and sealed with a PTFE film. The rotor was capped with a Vespel drive cap and quickly inserted into the pre-cooled DNP NMR probehead for measurement. The sample mass, dilution ratio of KBr, and the volume of the radical solution are listed in Table S1.

All DNP NMR experiments were performed on a 14.09 T AVANCE III HD spectrometer, corresponding to $^1$H Larmor frequency of 600 MHz, with a 395 GHz gyrotron microwave (MW) source and using a 3.2 mm triple resonance wide-bore probe. All experiments were performed at 12.5 kHz MAS frequency. A microwave source power of 11 W (at the source, equivalent to 110 mA collector current) was used for $^1$H-$^{13}$C DNP experiments. All MW on/off experiments were performed with a train of saturation pulses prior to a longitudinal relaxation delay followed by signal excitation. The characteristic build up time of the enhanced $^1$H polarization was measured via a saturation recovery experiment. The $^1$H enhancement ratios with microwave on and off are listed in Table S1. $^1$H-$^{13}$C CP experiments were performed with 90 – 100 % ramp on the $^1$H channel and 100 kHz $^1$H decoupling using swept-frequency two-pulse phase modulation (SW$_f$-TPPM) sequence[28]. The relaxation delay in the CP experiments varied between 4 - 7 s, with a CP contact time of 2 ms. $^1$H chemical shifts were referenced externally to the $^1$H and $^{13}$C (of CH$_2$ group) resonances of adamantane set at 1.8 and 38.6 ppm, respectively. Note for the LP30 +$^{13}$C$_3$ FEC sample, only 1 mg was used for measurement. The small sample amount is due to the small quantity of the $^{13}$C$_3$ FEC that was obtained from the synthesis.

## RESULTS AND DISCUSSIONS

**Electrochemistry.** The electrochemical performance of SiNWs cycled in LP30 and LP30 + FEC electrolytes is shown in Figure 1. In Figure 1a, discharge/charge (lithiaton/delithiation) capacities on the order of the theoretical capacity of Si (3579 mAh/g) were obtained for both LP30 and LP30 + FEC electrolytes during the first two cycles. However, over long-term cycling, obvious deviations are observed between LP30 and LP30 + FEC samples. At the 30th cycle, the LP30 sample exhibits only 55% capacity retention whereas, the LP30 + FEC sample retains 89% of the initial charge capacity (Figure 1b).

The voltage profile of SiNWs cycled in LP30 and LP30 + FEC during the first two cycles are similar (Figure 1a, left), indicating that both systems undergo similar structural transformations during the initial discharge/charge cycles. During the first discharge, the voltage quickly drops from the open circuit voltage (OCV) to 0.2 V with a small lithiation capacity, suggesting little SEI formation on the SiNWs from OCV to 0.2 V. The dQ/dV plot (Figure 1a, right) reveals the reduction process of FEC at 1.2 V and the reduction of EC at 0.8V during the 1st cycle. A flat discharge profile is then observed at approximately 0.1 V, which corresponds to the conversion of crystalline Si to amorphous lithium silicide (a-Li$_x$Si).[29] Further lithiation results in the formation of crystalline Li$_{15}$Si$_4$ (c-Li$_{15}$Si$_4$), which is manifested as a characteristic process at approximately 0.4 V in the charge voltage curve. The 0.4 V process corresponds to the delithiation of c-Li$_{15}$Si$_4$ and the formation of amorphous silicon (a-Si).[30] The c-Li$_{15}$Si$_4$ phase is highly reactive and induces severe electrolyte decomposition.[31] For the second discharge, the SiNWs show a voltage profile characteristic of a-Si. The two sloping processes at about 0.25 V and 0.10 V correspond to the lithiation stages (a-Si + x Li$^+$ → a-Li$_x$Si, where x is approximately 2.5 and 3.5 for the processes at 0.25 V and 0.1 V, respectively).[22] On the 30$^{th}$ cycle, the voltage profiles of the SiNWs cycled in LP30 or LP30 + FEC diverge (Figure 1a). Here, the SiNWs cycled in the presence of FEC maintain a voltage profile that is similar to that of the second cycle. In contrast, the onset of lithiation in the 30$^{th}$ cycle of the SiNWs is lower in LP30 than LP30+FEC sample as seen more clearly in the dQ/dV plot (Figure 1a, 30$^{th}$ cycle), suggesting a larger internal resistance inside the cell. This can be attributed to the formation of a more resistive SEI and increased electrode tortuosity, which limits Li ion diffusion through the bulk of the electrode and ultimately decreases lithiation capacity unless extremely low currents are used.[23,32]

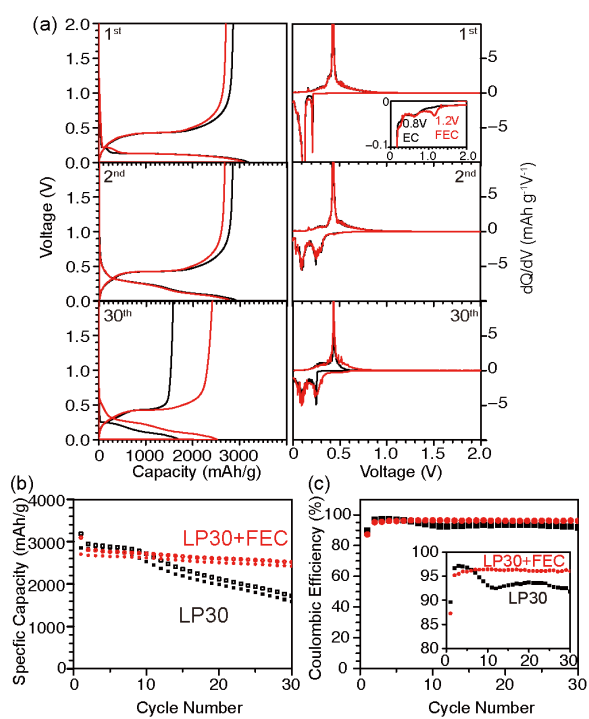

Figure 1. Electrochemical performance of SiNWs half-cells cycled with LP30 electrolyte (1 M LiPF$_6$ in EC/DMC=50/50, v/v, black), and LP30 with 10 vol % FEC (LP30 + FEC, red) electrolytes at the rate of C/30 (120 mAg$^{-1}$) between 0.001-2 V at room temperature. (a) the galvanostatic charge-discharge profiles and the corresponding dQ/dV plots of SiNWs cycled in LP30 and LP30 + FEC in the 1$^{st}$, 2$^{nd}$, and 30$^{th}$ cycles, (b) the cycling stability and (c) the coulombic efficiency for LP30 (black squares) and LP30 + FEC (red dots). The open dots/squares denote the discharge/lithiation capacity and the filled ones denote the charge/delithiation capacity.

The coulombic efficiency (CE, defined as delithiation capacity versus lithiation capacity) of SiNWs in LP30 and LP30 + FEC is compared in Figure 1c. During the first five cycles, the FEC sample shows a slightly lower CE than the LP30 sample, which may be due to the preferential decomposition of FEC over EC. From cycles 5 - 30, the average CE of LP30 + FEC is 96.2%, which is an improvement over LP30 alone (average CE 94.0%) but is still much lower than required for a practical cell, emphasizing the need for further understanding of the chemistries that influence CE.

**Soluble degradation products as measured by solution NMR.** Electrolytes from cells after the 1$^{st}$ and 30$^{th}$ cycles were compared with pristine electrolytes using $^1$H solution NMR spectroscopy (Figure 2) and a series of two dimensional (2D) correlation experiments. Several new $^1$H NMR signals were detected in the cycled LP30 electrolyte between 3 - 5 ppm (Figure 2a) that are not present in the FEC-containing samples (Figure 2b,c) indicating that very different soluble breakdown products are formed. In the LP30 samples, these include an intense singlet at 4.30 ppm that appears after the 1$^{st}$ cycle (yellow shading) and several multiplets (blue shading, labeled from a-d) at 4.19, 3.62, 3.52 and 3.4 – 3.1 ppm.

The singlet at 4.30 ppm is assigned to lithium ethylene dicarbonate (LEDC),[33] which is supported by 2D $^1$H-$^{13}$C heteronuclear single quantum correlation (HSQC) and $^1$H-$^{13}$C heteronuclear multiple bond correlation (HMBC) experiments performed on cycled LP30 electrolytes extracted from Li symmetric cells (Figure S3, experimental details described in SI) as well as previously reported DFT shift calculations.[34] LEDC is a decomposition product of EC, which is formed via a ring-opening reduction of EC, followed by a dimerization and the elimination of ethylene gas (Scheme 2.1).[35] Interestingly, LEDC disappears by the 30$^{th}$ cycle, suggesting that LEDC is a metastable species that decomposes upon further cycling. Decomposition of LEDC is consistent with theoretical predictions that indicate that LEDC is thermodynamically unstable on contact with the lithiated silicides.[36,37]

The multiplets labeled a-d are assigned to oligomers comprising different linear polyethylene oxide (PEO) species, i.e., R-OC*H*$_2$C*H*$_2$O- /R'-OC*H*$_3$ groups. Similar PEO species or oligomers have been previously detected by mass spectroscopies.[38–40] We find that these oligomers are extremely sensitive to trace amounts of water, with noticeable changes in $^1$H NMR peak positions as well as the emergence of new signals as moisture permeates the nominally airtight NMR tube (Figure S2).

The $^1$H NMR spectra of the LP30 + FEC electrolyte after 30 cycles exhibits three distinct new sets of resonances in the 9.5 - 9.7 ppm region (three singlets), a singlet at 7.77 ppm, and a cluster of multiplets at approximately 5.0 - 6.2 ppm (Figure 2b, shaded red and labeled x, y, and z, respectively), which are all absent in the LP30 sample (see Figure S1 for the expanded $^1$H NMR spectra). These three sets of resonances are also observed in the cycled LP30 + $^{13}$C$_3$ FEC with further splitting of the peaks resulting from the $^{13}$C labeling. Assignment of the species present in the FEC-containing samples was facilitated by a combination of 2D correlation NMR spectroscopy and J-coupling pattern analysis of the $^{13}$C-labeled sample (vide infra). By contrast, the 1D $^1$H NMR spectra of the LP30 sample only shows a small singlet at 8.42 ppm in this spectral region after 30 cycles, which can be assigned to lithium formate on the basis of its unique chemical shift.[41] Lithium formate can form via reduction of CO$_2$ and proton abstraction from other organic molecules in solution (Scheme 2.3).[34] The resonances seen in 5-10 ppm region in $^1$H NMR indicate that distinct chemical species are formed in the presence of the FEC additive that are not formed in LP30 alone.

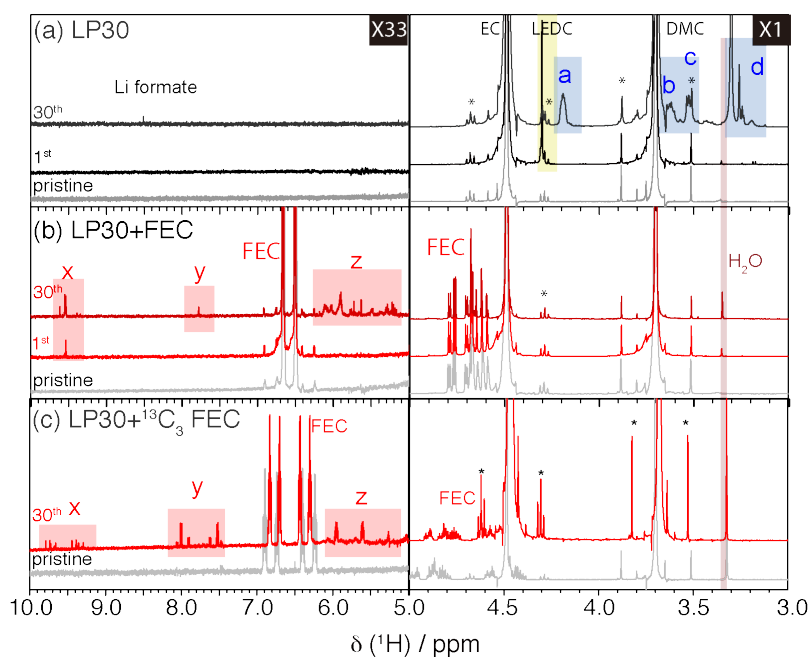

Figure 2. $^1$H solution NMR of (a) LP30 and (b) LP30 + FEC (c) LP30+$^{13}C_3$ FEC (LP30 + 10 vol% $^{13}C_3$ FEC) before cycling (pristine), and after the 1st and 30th cycles. The 5 – 10 ppm region is enlarged by 33 times compared to the 3 – 5 ppm region. $^{13}$C satellites are marked with an asterisk. All spectra were measured with a magnetic field strength of 9.4 T, except for the 30th cycle LP30+$^{13}C_3$ FEC sample that was measured at 11.7 T.

In addition to chemical composition, $^1$H NMR data also provide information on the relative populations of electrolyte breakdown products. In Figure 2, the $^1$H NMR resonances in the 5 – 10 ppm region are magnified by a factor of 33, over the 3 – 5 ppm region. The intense peaks of the degradation product seen in the LP30 sample (peaks a-d in Figure 2a) in the latter region suggest that more soluble oligomers are present in the electrolyte. In contrast, the $^1$H NMR peak intensities associated with the decomposition products found in the LP30 + FEC sample are significantly weaker than those in the LP30 sample, suggesting less soluble SEI are formed in the presence of FEC.

Two-dimensional (2D) correlation NMR spectroscopy experiments were then performed with $^{13}C_3$ FEC, in order to carry out a more in-depth characterization of the structure of the decomposition products, $^{13}C_3$ FEC being synthesized as described in the experimental section using a modification of a published route.[42] The HSQC spectrum of the LP30 + $^{13}C_3$ FEC samples (Figure 3a) shows two cross peaks between the $^1$H NMR signal at 9.53, 9.34 ppm and the $^{13}$C NMR signals at 188.5, 195.7 ppm, which are consistent with a terminal aldehyde/vinoxyl species, **_HC_**(=O)-R. The HMBC spectrum (Figure S7) shows that there are multiple vinoxyl oligomers, with the **_HC_**(=O) groups being bound to ethylene oxide (-**_C_**H$_2$O-) carbons with $^{13}$C chemical shifts of either 68.3, 71.9 or 73.8 ppm (all these $^{13}$C shifts are consistent with a formula such as HC(=O)CH$_2$OR). The carbon connectivity of this structure is further supported by $^{13}$C-$^{13}$C correlation spectroscopy (COSY, Figure 3b), which shows a cross peak between the vinoxyl carbons at 204.0 ppm and the ethyleneoxide carbons (-**_C_**H$_2$O-) around 68-73 ppm (see Table S3 for a summary of all 2D correlation peaks).

In addition to the vinoxyl species, other soluble components are also observed in the 2D NMR spectra. According to the HSQC spectrum in Figure 3a, the $^1$H NMR peak at 7.77 ppm (y) belongs to a proton directly bound to a sp$^2$ hybridized carbon (as indicated by the $^{13}$C chemical shift of 132.9 ppm). In addition, the corresponding $^1$H-$^{13}$C HMBC spectrum (Figure S7) shows that the proton giving rise to y is also 2 – 3 bonds away from a carbonate group, since a $^{13}$C cross peak is observed at 153.8 ppm. Thus y likely originates from a highly symmetric decomposition product of FEC, e.g., either vinylene carbonate (VC) or lithium vinyl dicarbonate (LVDC), both of which contain the chemical fragment O**_CH_**=CHR. Note that in the non-labeled sample (or under $^{13}$C decoupling), a singlet at 7.77 ppm is observed, whereas in the $^{13}C_3$ FEC sample, a distinct pattern of multiplets is observed (Figure 2b and 2c, region y). Of note, the splitting pattern observed in the $^{13}$C labeled sample contains further information that allows us to unravel the environments that give rise to the peaks in region y.

**Assignment of VC from Analysis of the J-coupling:** The experimental $^1$H NMR spectrum in region y of the cycled LP30 +$^{13}C_3$ FEC sample is compared with the simulated $^{13}C_2$ VC spectrum as shown in Figure 4. In order to simulate the proton-splitting pattern of $^{13}C_2$ VC, we first extracted various J coupling constants ($^1J_{CH}$, $^2J_{CH}$, $^1J_{CC}$, $^3J_{HH}$, $^3J_{CH}$) from the $^1$H NMR spectrum of natural abundance VC, which is also associated with a $^1$H shift of 7.77 ppm (Figure S9). The experimental J coupling constants are listed on the inset of Figure 4a. These J coupling constants were then used to simulate the $^1$H NMR spectrum of $^{13}C_2$ VC, by considering the four spin system, AA'XX' (Figure 4b). Here, we simplified the simulation by omitting $^3J_{CH}$ and, as a result, neglected the coupling to the carbonate group even though this sample originated from $^{13}C_3$-labeled FEC. (We note that in an AA'XX' system, each proton (A) is coupled to a different carbon (X), which gives second-order multiplets because $^1J_{CH}$ is different from $^2J_{CH}$). The appearance of the spectrum is defined by the four coupling constants ($^1J_{CH}$, $^2J_{CH}$, $^1J_{CC}$ and $^3J_{HH}$).

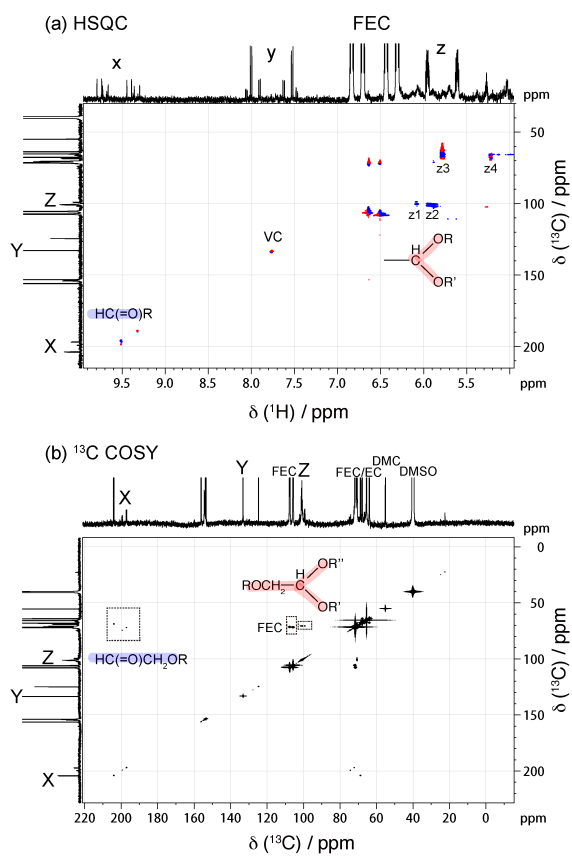

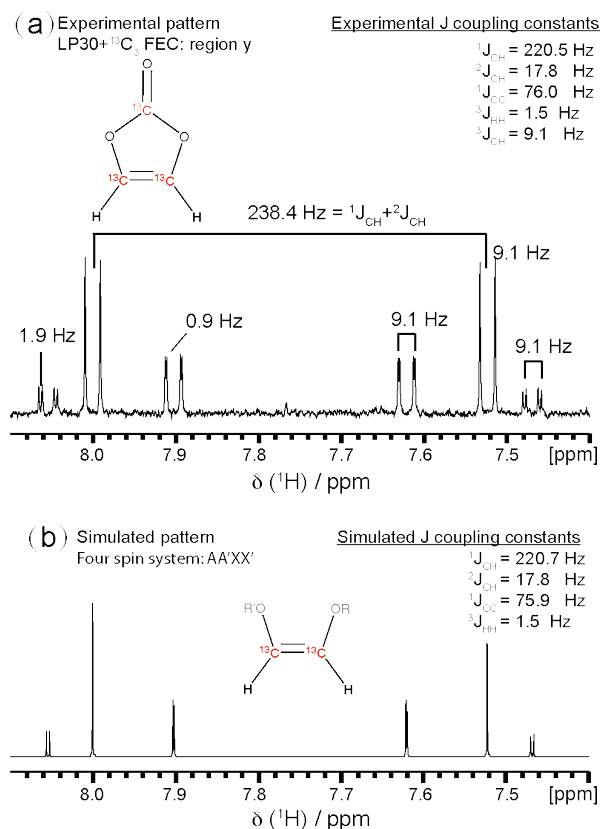

Figure 3. 2D solution NMR spectra of the LP30 + $^{13}C_3$ FEC electrolyte after 30th cycles. (a) $^1$H-$^{13}$C HSQC with $^{13}$C decoupling, blue and red represent positive and negative peaks, respectively. (b) $^{13}$C-$^{13}$C COSY spectra. The off-diagonal peaks are marked with dashed squares. Possible structures are given next to the corresponding peaks; species containing aldehyde terminal groups are shaded in blue and the cross-linking units are shaded in red.

Figure 4. Multiplet pattern of cycled LP30 + $^{13}C_3$ FEC in region y, (a) experimental pattern (the inset on the upper right-hand corner is the J-coupling constants of VC obtained from the $^1$H NMR spectrum of VC as illustrated in Figure S9); (b) simulated pattern of a four-spin system AA'XX' (cis-H-CR=CR-H) with the J-coupling constants used in the simulation listed on the upper right corner.

The peak position and its intensity can be calculated as described by Pople.[43–45] A least squares minimization was carried out to adjust the J coupling constants in order to match the experimental pattern. Figure 4b shows the simulated $^{13}C_2$ VC pattern with the corresponding J coupling constants. The simulation provides an excellent match to the experimental spectrum with the exception that the $^3J_{CH}$ coupling (9.1 Hz) is omitted since we only considered a four spin system; including this would lead to the observed 9.1 Hz splitting of all of the peaks. In contrast, LVDC, which contains carbonate groups on both ends of the molecule, will have a more complex $J_{CH}$ multiplet pattern: $^4J_{CH}$ (<10 Hz),[46] in addition to $^3J_{CH}$ couplings will exist, leading to additional splitting of the $^1$H signals (a doublet of doublets). Therefore, we can assign region y to VC and not LVDC. LVDC is also excluded on the basis of the measured $^3J_{HH}$ value (1.5 Hz), this coupling constant likely arising from a cis conformation (as in VC) rather than a trans one, which would be associated with a larger $^3J_{HH}$ value.[47]

**Assignment of Branched Oligomers and Vinoxyl Species:** The third region, labeled z, in the $^1$H NMR of the cycled LP30 + $^{13}C_3$ FEC sample shows multiplets at 6.07, 5.88, 5.78, and 5.20 ppm (which are labeled as z1, z2, z3 and z4 in Figure 3a). From the peaks observed in the HSQC spectrum, the $^1$H resonances z1 and z2 are connected to carbon resonances at 99.1 and 100.6 ppm, respectively. These resonance can be assigned to a protonated carbon with two oxygen groups attached (-**CH**(OR)$_2$) on the basis of its chemical shift (and its similarity to the shifts found in polysaccharides with similar local environments[48]). In the $^{13}$C-$^{13}$C COSY (Figure 3b) the branched carbons at 100.3 and 98.8 ppm are directly bound to the ethylene oxide carbon at 70.2 ppm (Figure S8 for details), suggesting a motif structure: RO**C**H$_2$**C**H(OR)$_2$. This key observation is indicative of the formation of branched oligomers in the FEC-containing electrolyte, which appears after prolonged cycling.

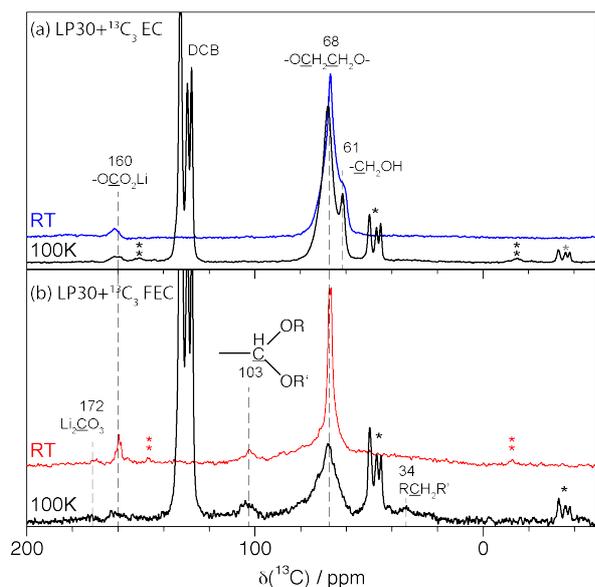

Figure 5. $^1$H-$^{13}$C CP NMR spectra of SiNWs cycled in LP30 with 25 vol %$^{13}$C$_3$ EC (LP30 + $^{13}$C$_3$ EC) (a) and LP30 with 10 vol% $^{13}$C$_3$ FEC (LP30 + $^{13}$C$_3$ FEC) (b) electrolytes, for 30 cycles. The RT spectra were measured at room temperature by conventional ssNMR, while the 100 K spectra were measured using DNP NMR with the microwaves turned on. Ortho-dichlorobenzene (DCB) was used as a radical solvent in the DNP experiments and its isotropic resonances are labeled with "DCB"; the spinning sidebands of all the resonances are marked with asterisks. Possible structures are given next to the various isotropic resonances where R represents CH/CH$_2$/CH$_3$ groups.

Similarly, structural assignments for $^1$H NMR at z3 and z4 can be made. $^1$H resonance at z3 is connected to a $^{13}$C resonance at 65.8 ppm in the HSQC spectrum and the protons are 2-3 bonds away from carbons with $^{13}$C resonances at 66.4 and 153.2 ppm in the HMBC spectrum. These resonances can be assigned to a branched carbon near a carbonate group and two ethylene oxide groups (ROCOO**CH**(CH$_2$OR')$_2$) based on the their chemical shifts. The proton with $^1$H resonance at z4 is bound to a carbon at 66.7 ppm. In HMBC, z4 is further bonded to $^{13}$C at 68.2 and 203.3 ppm. The chemical structure of z4 can be assigned to the methylene units in the vinoxyl species (HC(=O)**CH$_2$**OCH$_2$R), which confirms the identification of the vinoxyl species in region x (see Table S3 for a summary of all the correlation peaks in solution NMR spectra and possible assignments for the LP30+$^{13}$C$_3$ FEC sample).

**$^{13}$C ssNMR and DNP NMR Detection of the SEI:** Characterization of the insoluble species in the SEI was carried out with $^{13}$C$_3$-labeled EC, $^{13}$C$_3$-labeled FEC electrolyte, and electrolyte/additive formulations (see Table 1 for electrolyte formulation) using a combination of ssNMR and DNP NMR spectroscopies. Cycled SiNWs were extracted from cells without rinsing and were dried under vacuum overnight to remove EC/DMC before measurement. Figure 5 shows a comparison of the $^1$H-$^{13}$C cross-polarization (CP) NMR spectra measured at room temperature (RT) using conventional ssNMR and the spectra acquired with DNP NMR at 100 K. All DNP spectra show intense DCB solvent peaks between 120 - 140 ppm with the corresponding spinning sidebands at 40-55 ppm (Figure 5) due to the addition of the DNP biradical solution (4 mM TEKPol in DCB). Apart from these solvent peaks, there is no obvious difference in the species detected via ssNMR and DNP NMR, suggesting that the biradical solution has not altered the chemical structure of the SEI. Moreover, the sensitivity provided by low temperature DNP is obvious - the room temperature (RT) spectrum of the LP30 + $^{13}$C$_3$ FEC sample took approximately 20 h, whereas, under DNP conditions (100 K), a similar signal-to-noise ratio spectrum was achieved within 1.3 h – and allowed characterization of the SEI via $^1$H-$^{13}$C heteronuclear correlation (HETCOR) experiments (Figure 6).

**LP30:** The $^{13}$C NMR spectrum of LP30 containing 25 vol % enriched EC (LP30 + $^{13}$C$_3$ EC, Figure 5a) is dominated by a broad peak at 68 ppm with a shoulder at 61 ppm. In addition, a semicarbonate resonance at 160 ppm is observed. The shoulder at 61 ppm becomes sharper in the spectrum acquired with DNP at 100 K, likely due to reduced dynamics of the organic SEI species at lower temperatures, as observed in our previous DNP experiments performed on graphene electrodes.[49] The broad $^{13}$C resonance at 68 ppm (labeled C1) is correlated to a proton at 4.5 ppm in the HETCOR (Figure 6a), allowing C1 to be assigned to either the carbon in ethylene oxide (-**C**H$_2$**C**H$_2$O-) or to residual EC. The shoulder at 61 ppm in the $^{13}$C ssNMR spectrum (labeled C2) shows a one-bond correlation to a proton resonance at approximately 3.75 ppm, and is assigned to an ethylene oxide carbon with a terminal alcohol (R**C**H$_2$OH).[34] Three other local maxima are observed in the HETCOR spectra between C1 and C2 (numbered 3, 4 and 5 in Figure S13), the different shifts possibly being a result of different PEO chain lengths and/or a variation in terminal groups of polyethylene oxide. The intensity of aliphatic carbons in both the conventional and DNP $^{13}$C NMR spectra is low, which suggests that the SEI formed here primarily consists of polyethylene oxide that contains few aliphatic units.

**LP30 + FEC:** Similar carbon environments at 68 ppm and 160 ppm are observed in the spectrum of LP30 with 10 vol % of $^{13}$C-enriched FEC (LP30 +$^{13}$C$_3$ FEC, Figure 5b), along with two new peaks: a $^{13}$C resonance at 103 ppm that is present in both the ssNMR and DNP, as well as a weak resonance at approximately 34 ppm that is much more clearly resolved in the DNP spectrum. The main peak at 68 ppm has a different peakshape compared to that observed in the LP30 + $^{13}$C$_3$ EC spectrum, and no shoulder at 61 ppm is observed. The signal at 68 ppm in the RT spectrum is broadened near the baseline, possibly indicating that two peaks are superimposed in this region. We hypothesize that the sharp peak at 68 ppm is due to residual EC. The broader component of the peak as well as the $^{13}$C resonance at 68 ppm is consistent with a distribution of different ethylene oxide environments (-CH$_2$CH$_2$O-).[34] In contrast to the LP30 + $^{13}$C$_3$ EC sample, the SEI signal at 68 ppm becomes broader when measured at 100 K (Figure 5b). Here, spectral broadening may be a result of sample

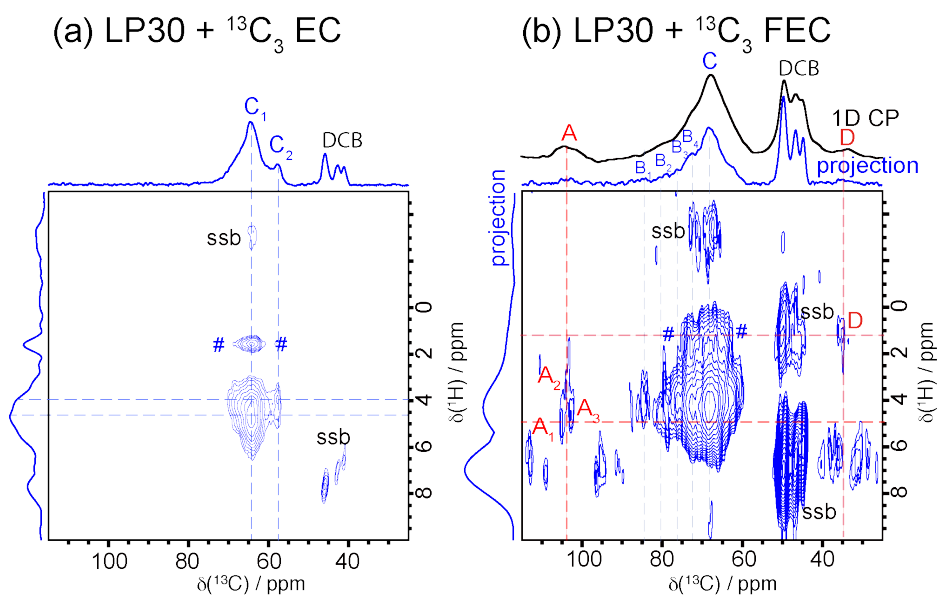

Figure 6 2D $^1$H-$^{13}$C heteronuclear correlation (HETCOR) DNP-NMR of SiNWs cycled in LP30+$^{13}$C$_3$ EC electrolyte (a) and LP30 + $^{13}$C$_3$ FEC electrolyte (b) for 30 cycles. Spinning sidebands arising from the DCB solvent are marked with ssb; artifact peaks marked with # are due to spin locking along the effective field proton decoupling and they appear at the $^1$H carrier frequency.[50] Full spectra and additional experimental details can be found in the SI.

heterogeneity, leading to a wider distribution of chemical shifts at low temperature.[51]

The HETCOR spectrum of the FEC-containing sample (Figure 6b) has three extra peaks at 103, 75-85, and 34 ppm, labeled A, B, and D, respectively, which are absent in the LP30 + $^{13}$C$_3$ EC sample. The broad peak, B, is correlated to a $^1$H peak at around 4.5-5.0 ppm. At least four components (B1 – B4) can be resolved, which are assigned to branched ethylene oxide units (-CHRO-; see Scheme 3) with different substituting groups or chain lengths. The $^{13}$C resonance at 34 ppm (peak D) is correlated a $^1$H resonance at 1.3 ppm, and is assigned to RCH$_2$R' units. Both the aliphatic units (peak D) and the series of resonances round 80 ppm (peak B) are only present in the FEC containing sample and imply that the structure that originates from FEC is more complex than a simple linear ethylene oxide polymer.

In the HETCOR spectrum, the peak at 103 ppm (peak A) is bound to protons that resonate between 4.2 - 5.2 ppm, and can be assigned to a protonated carbonate bound to two oxygens based on prior DFT shift calculations of proposed cross-linked VC polymers[7] (see Scheme 3). This unique chemical shift is consistent with branched structures and was observed in the spectrum of chemically reduced FEC.[7] The peak at 103 ppm is not due to residual FEC as no FEC was detected by $^{19}$F ssNMR (Figure S10). Interestingly, similar resonances were also detected in solution NMR of cycled LP30 + FEC electrolyte (resonance z1 and z2: the proton at 5.78 ppm is bonded to a carbon at 100 ppm in HSQC, Figure 3a), suggesting that these might be the precursors that eventually form the insoluble, higher molecular weight SEI polymers as will be discussed later. Similar cross-linking units are present as glycosidic linkages in natural polysaccharides and also exhibit $^{13}$C resonances close to 100 ppm.[48]

## DISCUSSION

Different electrolytes with $^{13}$C$_3$ enriched EC and/or $^{13}$C$_3$ enriched FEC were cycled in SiNW half-cells to study the organic electrolyte degradation products. FEC-containing electrolytes display an obvious improvement on the cycle life of Si anode compared to electrolytes that do not contain FEC. The SiNWs cells were stopped at the delithiated state after the 1$^{st}$ and 30$^{th}$ cycle for ex-situ NMR analysis. The cycled electrolytes were examined by solution NMR and the electrodes by solid-state and DNP NMR spectroscopies. Both soluble and insoluble chemical structures detected in LP30 samples with and without FEC are listed in Schemes 2-3 along with possible formation reactions.

For the LP30 sample, soluble products such as LEDC, lithium formate and PEO-type oligomers are detected in the cycled electrolyte and their respective formation pathways are shown in Scheme 2. The insoluble SEI that forms from LP30 mainly consists of ethylene oxides (-CH$_2$CH$_2$O-), ethylene oxides with hydroxide terminal units (RCH$_2$OH) and carbonate (-OCOO-) units. The presence of these ethylene oxide species is consistent with the PEO-type and lithium alkyl carbonate polymers formed from EC that were reported by Shkrob.[52]

Previous works on EC decomposition[53–55] suggest that EC can undergo one-electron ring-opening reduction to form a lithium alkyl carbonate anion radical (Scheme 2.1). The radical can then dimerize and form LEDC with concurrent loss of ethylene gas, which has been previously

**Schemes 2-3.** Possible reaction schemes consistent with the chemical signatures detected by solution and solid-state NMR for (2) the LP30 sample: (2.1) reduction of ethylene carbonate[52,56], (2.2) anionic polymerization of EC, (2.3) formation of lithium formate, (2.4) two possible reactions for the formation of ROH[54,57]. (3) LP30 + FEC sample: (3.1) reduction of fluoethylene carbonate; (3.2, 3.3) examples of reactions between VC and vinoxyl radicals as the initial steps for radical polymerization to form poly(VC); (3.4, 3.5) reactions between alkene termination and vinoxyl radicals; (3.6) possible reactions between vinoxyl radical B and aldehyde species forming the vinoxyl units of the type detected by solution NMR; (3.7) reaction between secondary radical formed in (3.5) and aldehyde species, forming branched structure found in solution NMR; (3.8) possible reaction between alkene termination with reduced EC intermediate. The aldehyde terminal units and vinoxyl species are all shaded in blue. R/R' groups are organic fragments. The cross-linking units, branched ethylene oxide and aliphatic carbon are shaded in red, yellow and green, respectively, along with their corresponding $^{13}C$ chemical shifts marked.

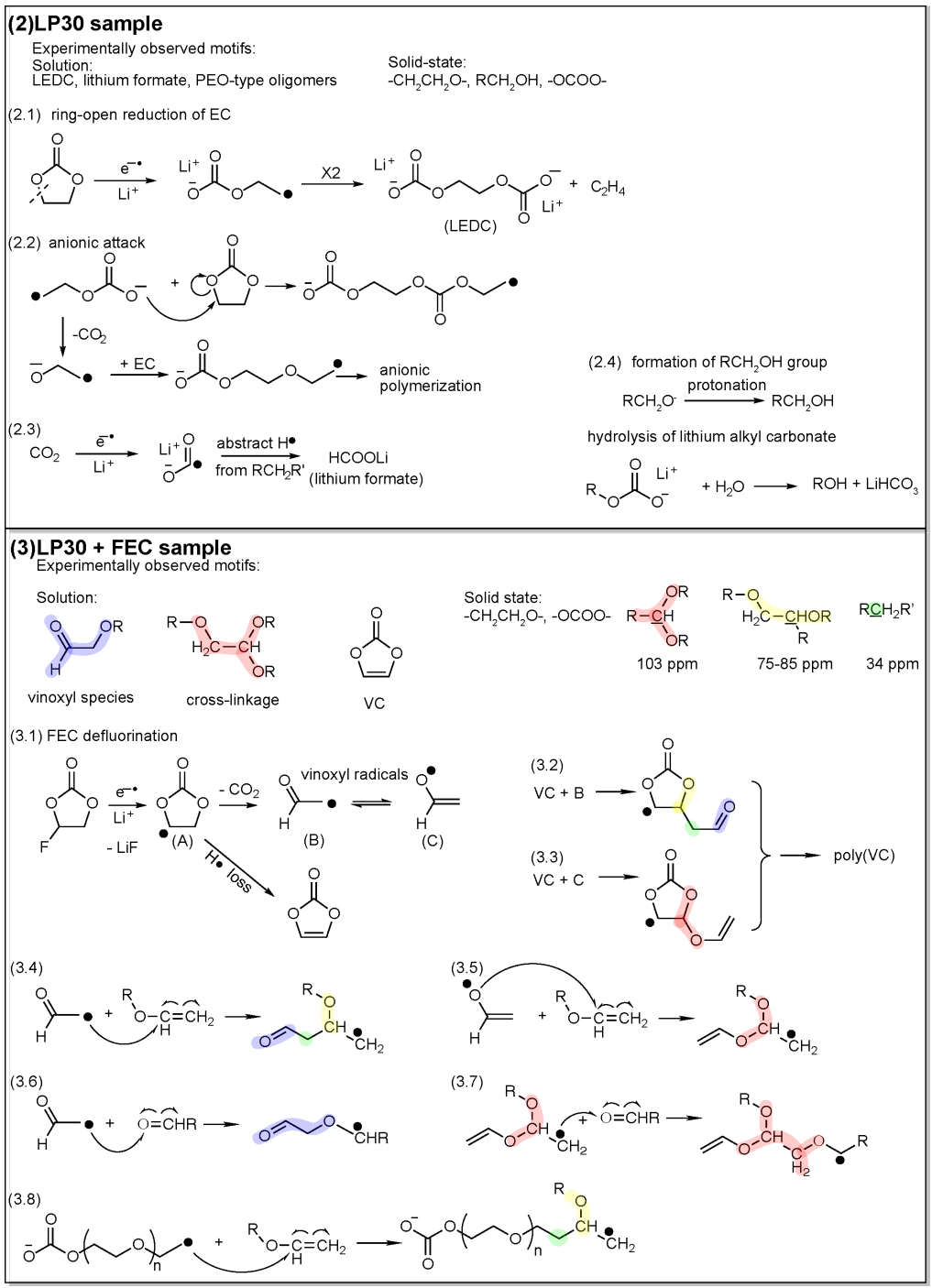

detected by GC-MS (Scheme 2.1).[58] To form oligomer or polymers, the alkyl carbonate radical anion can elongate the chain via a nucleophilic attack on the EC. Alternatively, it can lose $CO_2$, forming an ethylene oxide radical anion ($•CH_2CH_2O^-$).[56] This radical anion can also attack EC to initiate anionic polymerization (Scheme 2.2).[56] In either case, the resulting polymer will primarily consist of linear carbonates and ethylene oxide units which are consistent with the ssNMR data in this study as well as those published in previous report.[23] During these processes, lithium formate can be formed by the reduction of $CO_2$, resulting in a lithium carbon dioxide radical with subsequent hydrogen abstraction from other species in the solution (Scheme 2.3).[7]

Short-chain lithium alkyl carbonates such as LEDC are highly soluble and thus are unable to protect the Si anode, as they are easily detected in the cycled electrolyte. LEDC is only transiently detected in the 1st cycle electrolyte and not in the long-term cycled electrolyte, confirming its instability.[36] Significant amounts of PEO-type oligomers are present in the electrolyte after prolonged cycling, which suggests that the SEI derived from LP30 is highly soluble. The existence of these soluble degradation products also suggest that they are unable to adhere to the Si surface and thus prevent further electrolyte decomposition. The instability of the SEI formed from LP30 contributes to the irreversible consumption of lithium and it agrees well with the capacity fading observed in the electrochemistry of the LP30 sample (Figure 1b).

The insoluble polymeric SEI formed from LP30 contains similar chemical units to those detected in the solution NMR. Note that significant amounts of hydroxide-terminated groups ($RCH_2OH$) are detected in the ssNMR of the LP30 sample, which is consistent with previous study.[23] Hydroxide terminal units can be formed via protonation of the alkoxide ($RCH_2O^-$) or by the hydrolysis of lithium alkyl carbonate (Scheme 2.4).[54,57] Although the $^{13}C$ CP NMR is not quantitative, the long contact time used in the experiments here (1-2 ms) ensures a homogenized $^1H$ polarization transfer throughout the molecule, allowing a semi-quantitative comparison. The higher percentage of hydroxide-terminated carbon versus ethylene oxide carbon in the LP30 sample suggests a higher population of polymers with shorter chain lengths compared to highly polymerized PEO species. Such short-chain polymers/oligomers are likely to be chemically similar to the oligomers detected in the cycled electrolyte (peak a-d in Figure 1a). These short-chain polymers derived from EC do not appear to be able to form a stable SEI on the Si anode and help to prevent further electrolyte breakdown.

Very different chemical motifs were detected in the solution and solid-state NMR of the FEC-containing sample (Scheme 3, top). Specifically, minor amounts of vinoxyl species, VC, and a possible cross-linking site are present as the soluble products in the cycled electrolyte. Note that PEO-type oligomers were not detected by solution NMR in the FEC sample. The absence of soluble PEO-type species suggests that FEC can effectively suppress the formation of soluble oligomers formed in the LP30 sample.

Vinoxyl species are present in both the 1st and 30th cycle, indicating that they form during the initial stage of FEC decomposition. In contrast, VC and branched oligomers are only found in the 30th cycle of the FEC-containing sample. We speculate that VC may be highly reactive and thus, rapidly being consumed to form other species in the initial formation cycles. However, after long-term cycling, a stable SEI forms and VC begins to accumulate in the electrolyte. The conversion of FEC to VC that we observed is also consistent with the mechanistic studied performed by Balbuena and co-workers using density functional theory and ab initio molecular dynamics simulations methods.[59] The branched soluble oligomers detected in the cycled electrolyte appear to be similar to the species present in the insoluble portion of the SEI, but with shorter chain lengths (i.e. lower molecular weight species and higher solubility; they therefore saturate in the electrolyte and likely prevent further SEI dissolution). The soluble components may also serve as precursor for the insoluble SEI polymers. Another possibility is that the chemical structures of the soluble oligomers are different from the insoluble polymer as they are formed by different reactions. Certain pathways lead to short-chain oligomers, while other reactions form insoluble polymer.

The insoluble SEI products formed from the FEC-containing sample are consistent with ethylene oxides and carbonate species along with the minor structural features as follows: acetal carbons (with $^{13}C$ chemical shift at 103 ppm), branched ethylene oxides (with $^{13}C$ chemical shift at 75-85 ppm), and aliphatic carbons (with $^{13}C$ chemical shift at 34 ppm), which are shaded in red, yellow and green, respectively (Scheme 3, top). The observation of ethylene oxides and carbonate species is consistent with prior NMR,[7] XPS[14,15] and FTIR studies[10].

Scheme 3 summarizes the possible reduction reactions of FEC, which are based on the species detected in this study and prior experimental and theoretical work. First, FEC is defluorinated, forming an EC• radical (denoted as radical A) and LiF (Scheme 3.1), as proposed by Nie.[35] As fluorinated carbon species are not detected by solution or solid-state NMR (Figure S10), we suggest that FEC defluorinates prior to further reaction. At this stage, the formed EC radical can abstract hydrogen from other species in solution and convert back to EC. Alternatively, the EC radicals can disproportionate to form VC and EC. Experiments using mass spectrometry in conjunction with additional NMR to compare decomposition products using unlabeled and $^{13}C$-labeled EC and FEC are currently underway in our laboratory to determine which reaction pathway is occurring and will be reported in a future study. If the EC radicals disproportionate to form VC and EC in the LP30 + $^{13}C_3$ FEC sample, the $^{13}C_3$-labeled EC that is generated can be reduced as suggested in Scheme 2 and subsequently contribute to the PEO-type signal that is detected in $^{13}C$ ssNMR (Figure 5a).

Radical A is identical to the radical that results from EC via H abstraction. However, although its existence has been proposed, such cyclic EC radicals have not been experimentally observed in the absence of FEC or VC, even under cryogenic conditions (77 K) during the irradiation of EC.[52] The inability of EC to form such an EC radical may be one explanation for the difference in the decomposition products seen with EC and FEC. Once formed, radical A can lose hydrogen to form VC, otherwise, radical A can lose $CO_2$, forming the vinoxyl radicals as shown in Scheme 3.1.

Due to resonance, there are two forms of the vinoxyl radicals: one with the radical center on the carbon ($CH(=O)CH_2•$, radical B) and the other with the radical center on the oxygen ($•OCH=CH_2$, radical C). While the vinoxyl radicals have not been directly detected in this work, such radicals have been observed in a radiolysis experiment on FEC and were proposed to initiate the formation of highly cross-linked polymer.[18] Because our NMR results only revealed stable vinoxyl species instead of unstable vinoxyl radicals, we now propose possible reaction schemes that result in the formation of vinoxyl species as well as some branched units with predicted chemical shifts similar to those observed experimentally (Scheme 3.2 - 3.8). Vinoxyl species can, for example, be formed by the vinoxyl radicals (either B or C) attacking the $sp^2$ hybridized carbon in VC. When radical B reacts with VC (Scheme 3.2), it can form a structure that contains an aldehyde terminal group, an aliphatic carbon, and a branched ethylene oxide (shaded in blue, green and yellow, respectively, with corresponding NMR parameters given). If VC reacts with radical C (Scheme 3.3), the radical will be transformed into a stabilized carbon radical, which has a branched acetal carbon (shaded in red) and an alkene termination. These newly formed radicals (in the form of RCH• R') can abstract H from other species in solution to stabilize themselves (forming $RCH_2R'$). Alternatively, these secondary radicals can then further react with the vinyl group in VC to form poly(VC).[7,37,58]

Note that neither alkene units ($^{13}C$ shifts at 120-140 ppm) nor aldehyde carbon ($^{13}C$ shift at 200 ppm) are observed by ssNMR. We hypothesize that such terminations can be consumed by further reacting with the vinoxyl radical and resulting in chain elongation.[60] Such terminal groups would be present in very low quantities and therefore, below the detection limit of ssNMR. Furthermore, decomposition products that contain alkene terminal units can undergo reactions similar to that of VC (Scheme 3.4 and 3.5), forming a mixture of polymer products that is consistent with the cross-linked species detected in this study.

Interestingly, the chemical structures observed in solution can be rationalized by considering the radical attack of the aldehyde terminal group. When radical B reacts with a molecular with an aldehyde group (Scheme 3.6), a new radical containing the vinoxyl units ($CH(=O)CH_2OR$, shaded in blue) is formed. When the secondary radical formed from Scheme 3.5 attacks an aldehyde group (Scheme 3.7), a cross-link containing the acetal carbon forms ($ROCH_2CH(OR)_2$, shaded in red). These two chemical units are consistent with the soluble products identified by solution NMR (Figure 3). We speculate that the radical attack on the aldehyde group will lead to oligomers that have short-chain lengths and remain solubilized. In contrast, radical attack on the alkene terminal group is more likely to form higher molecular-weight polymers that are incorporated into the insoluble portion of the SEI.

The reduced EC intermediate (alkyl carbonate anion radical) can also react with the alkene carbon as illustrated in Scheme 3.8. The anion radical ($RCH_2•$) that forms from reduced EC can attack the alkene group and graft the PEO chain to the decomposition products of FEC. If it occurs, this reaction also consumes the anion radicals and reduces the possibility of anionic polymerization of EC. In Scheme 3.8, $RCH_2R'$ (shaded in green) could, in principle, originate from the decomposition of EC. To determine whether this reaction takes place, $^{13}C$ CP NMR of the SiNWs cycled in LP30 + $^{13}C_3$ EC + FEC (see Table 1 for electrolyte formulation) was performed. The resulting $^{13}C$ CP NMR spectrum shows an extra set of resonances that span the range of 15 – 40 ppm (Figure S11). The presence of additional $^{13}C$ NMR peaks in the region of 15 – 40 ppm strongly suggests that EC contributes to the formation of the aliphatic carbon signal and is consistent with the mechanism proposed in Scheme 3.8.

Although alkene termination is not directly observed in solid-state NMR, we speculate that $sp^2$ carbon/alkene termination is necessary to create the cross-linked polymer, and may play an important role in capacity retention in general. Recent reports indicate that novel additives, such as methylene ethylene carbonate that contain $sp^2$ hybridized carbons show promise for increasing capacity retention in LIBs.[61,62] The SEI formed in the presence of FEC clearly shows cross-linked species, whereas the SEI formed in the standard EC/DMC electrolyte mainly contains linear PEO-type polymers, providing a molecular rationale for the observed increase in capacity retention in LIBs when FEC additive is used. Similar cross-linking units are also present as glycosidic linkages in natural polysaccharides, (which exhibit similar $^{13}C$ resonances at approximately 100 ppm[48]) many of which have been successfully demonstrated as a binder for Si to improve capacity retention, further suggesting that this structural motif may impart stability to the SEI.[63–65]

The mechanical properties of the branched polymer derived from FEC may be more elastic, which can accommodate the volume expansion that occurs in Si during cycling. Additional experiments are required to determine whether the polymers formed from FEC differ in their Li ion conductivity than the linear PEO-like species formed from EC and whether the reduced overpotentials seen on cycling in the presence of FEC are due to a thinner SEI or to improved Li transport.

## *CONCLUSION*

Organic species in the SEI on SiNWs were characterized by solution and solid-state NMR to understand the role of FEC as an electrolyte additive in performance enhancement in LIBs. After long-term cycling, the standard EC/DMC electrolyte decomposes and forms a variety of soluble oligomers in addition to the transient formation of LEDC. The addition of FEC into the electrolyte allows the formation of a stable SEI and suppresses the decomposition of EC/DMC, resulting in increased coulombic efficiency after the first few cycles. The $^1H$ and $^{13}C$ NMR spectra provide compelling evidence for the defluorination of FEC to form soluble vinoxyl species ($HCOCH_2OR$) and VC. Importantly, we emphasize that we have conclusively shown that FEC converts to VC instead of LVDC by $^1H$ NMR using $^{13}C$-labeled FEC. Oligomers with characteristic peaks that can be assigned to protonated carbons bonded to two adjacent oxygen groups due to cross-linking units were also identified. These oligomeric precursors presumably react further to form insoluble polymeric species in the SEI, with similar cross-linking groups. Neither these cross-linking units nor the vinoxyl species are observed in the absence of the FEC additive.

The vinoxyl species are signatures for the formation of the vinoxyl radicals that are believed to initiate the polymerization that eventually results in a highly cross-linked network.[18] While the study of Shkrob et al. focused on the reduction products of FEC alone,[18] we too, detect similar vinoxyl species and cross-linking motifs when FEC is used as an additive in EC-containing electrolytes. Based on our NMR results, we find that the stepwise elimination of $CO_2$ results in a polymeric species that contains a mixture of aliphatic units ($^{13}C$ shifts at 34 ppm) and cross-linking motifs ($^{13}C$ shifts at 103 ppm) similar to poly(VC), with several regions of PEO-type structures. Overall, FEC breakdown products (e.g. increased population of cross-linking moieties) lead to a suppression of soluble, linear PEO-type polymeric products that occur in the standard cycled LP30 electrolyte.

We speculate that the formation of cross-linked polymers is key to the higher stability of SEI formed on Si in the presence of FEC, motivating studies with additives that may promote cross-linking. Further insight into the molecular nature of the SEI and the param-eters that impart stability offer the opportunity to tailor the SEI chemistry to maximize performance in LIBs.

## ASSOCIATED CONTENT

### Supporting Information

The Supporting Information is available free of charge on the ACS Publications website.

Further solution NMR ssNMR and DNP NMR data and experimental details (PDF)

## AUTHOR INFORMATION

### Corresponding Author

* cpg27@cam.ac.uk

### Notes

The authors declare no competing financial interest.

## ACKNOWLEDGMENT

The authors thank Prof S. Hoffmann for providing access to the CVD system to synthesize the SiNWs, and Drs. A. Michan and Dr. Z. Liu for helping with the electrochemistry. YJ thanks Cambridge Trust and Chinese Scholarship Council for PhD funding. Financial support from the Engineering and Physical Sciences Research Council (U.K.) under the Supergen consortium and Amorphous grant (EP/N001583/1) is acknowledged. LEM gratefully acknowledges financial support through a FP7 Marie Cure International Incoming Fellowship. ECM acknowledges financial support through a H2020 Marie Sklodowska Cure Individual Fellowship. GK acknowledges funding from the European Unions's Horizon 2020 research and innovation programme under Grant Agreement No. 696656. The DNP experiments were performed at the DNP MAS NMR Facility at the University of Nottingham, with thanks to the EPSRC for funding of pilot studies (EP/L022524/1).

## Table of Contents

NMR studies of electrolyte decomposition in batteries

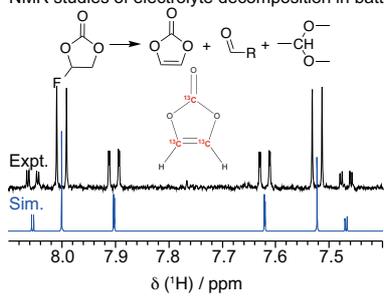